\documentclass[conference,10pt, a4paper,twocolumns,final]{IEEEtran}

\newlength{\flexwidth}
\usepackage{amsmath,amssymb,fixmath}
\usepackage{fixmath}
\usepackage{xcolor}
\usepackage{amsthm}
\usepackage{siunitx}
\usepackage{cuted}
\usepackage{multirow}
\usepackage{booktabs}
\usepackage{subcaption}
\usepackage[inline]{enumitem}
\usepackage{optidef}
\usepackage{graphicx}
\usepackage{lipsum}

\usepackage[norelsize,linesnumbered]{algorithm2e}
\makeatletter
\newcommand{\removelatexerror} {\let\@latex@error\@gobble}
\makeatother

\usepackage{tcolorbox}
\tcbuselibrary{many}

\newcommand{\superscript}[1]{^{\mathrm{#1}}}
\newcommand{\subscript}[1]{_{\mathrm{#1}}}

\newcommand{\ul}{\subscript{UL}}
\newcommand{\dl}{\subscript{DL}}
\newcommand{\cl}{\subscript{CL}}

\newtcbtheorem[]{alg}{Algorithm}{fonttitle=\bfseries}{alg}

\begin{document}
\title{Time-Energy-Constrained Closed-Loop FBL Communication for Dependable MEC}

\author{
	\IEEEauthorblockN{Bin~Han\IEEEauthorrefmark{1},~Yao~Zhu\IEEEauthorrefmark{2},~Anke~Schmeink\IEEEauthorrefmark{2},~and~Hans~D.~Schotten\IEEEauthorrefmark{1}\IEEEauthorrefmark{3}}
	\IEEEauthorblockA{
		\IEEEauthorrefmark{1}Division of Wireless Communications and Radio Positioning (WiCoN), University of Kaiserslautern\\%
		\IEEEauthorrefmark{2}ISEK Research Area, RWTH Aachen University\\%
		\IEEEauthorrefmark{3}Research Department Intelligent Networks, German Research Center of Artificial Intelligence (DFKI)%
	}
\thanks{Bin Han (binhan@eit.uni-kl.de) is the corresponding author.}
}

\markboth{Journal of \LaTeX\ Class Files,~Vol.~14, No.~8, August~2015}%
{Shell \MakeLowercase{\textit{et al.}}: Bare Demo of IEEEtran.cls for IEEE Journals}
%



\maketitle

\begin{abstract}
The deployment of multi-access edge computing (MEC) is paving the way towards pervasive intelligence in future 6G networks. This new paradigm also proposes emerging requirements of dependable communications, which goes beyond the ultra-reliable low latency communication (URLLC), focusing on the performance of a closed loop instead of that of an unidirectional link. This work studies the simple but efficient one-shot transmission scheme, investigating the closed-loop-reliability-optimal policy of blocklength allocation under stringent time and energy constraints.
\end{abstract}


%
\IEEEpeerreviewmaketitle

\section{Introduction}
\IEEEPARstart{A}{s} a key enabler of pervasive intelligence in future 6G mobile networks, multi-access edge computing (MEC) brings computation resources close to the mobile user devices, and therewith significantly reduces the latency of offloading intensive computing tasks onto the cloud~\cite{mach2017mobile}.

The performance of MEC highly depends on the quality of radio link~\cite{ren2020joint}, including the air interface latency, the radio link reliability, and the energy efficiency. Especially, for tolerance-critical applications such as remote control and factory automation, an ultra-reliable low latency communication (URLLC) between the user equipment (UE) and the MEC server is required~\cite{elayoubi20165g}. It is usually considered to use short codes in such systems for lower latency and better synchronization. However, diverging from the asymptotic regime and therefore having less capabilities of error correction, short codes generally exhibit worse SNR performance than long codes, leading to concerns in the link reliability. An emerging research interest on the finite blocklength (FBL) information theory has been therefore raised over the recent years. Following the pioneering work in \cite{polyanskiy2010channel}, various aspects of telecommunication technologies have been investigated in the FBL regime, including packet scheduling~\cite{xu2016energy}, power control~\cite{yang2015optimum}, radio resource management~\cite{she2017radio}, interference management~\cite{ozcan2013throughput}, energy harvesting~\cite{hu2019swipt}, etc.

Most existing studies over URLLC and FBL information theory are considering the uplink (UL) and downlink (DL) transmissions independent from each other, focusing on the analysis and optimization of an unidirectional link in open loop. However, in many dependability-critical applications such as factory automation and automated driving, it is the reliability and latency of closed-loop message exchange between the UE and the MEC server that matters.

In a recent previous work of ours~\cite{han2021clarq}, we have studied the automatic repeat request (ARQ) and hybrid ARQ (HARQ) schemes regarding the closed-loop reliability under a stringent closed-loop air interface latency constraint. Our work shows that the one-shot transmission scheme, which fully uses the entire time frame in one UL slot and one DL slot without any retransmission, outperforms any static ARQ/HARQ policy. It also proposes a dynamic ARQ mechanism, namely closed-loop ARQ (CLARQ), that dynamically re-splits the remaining time frame into an UL retransmission slot and a new DL slot. As a dynamic-programming generalization of the one-shot scheme, the CLARQ mechanism is proven superior over the one-shot scheme in both closed-loop reliability and energy efficiency, when specified to an arbitrarily fixed UL transmission power. However, as we have commented in \cite{han2021clarq}, it remains an open challenge to jointly optimize the transmission power and the blocklength of codewords.

As a breach and preliminary work towards joint optimization of power and blocklength in CLARQ, in this paper we investigate the energy-aware blocklength allocation in a simple but special case of CLARQ: the one-shot transmission scheme. Our study proves the existence of an unique optimal solution to this problem under some weak assumptions. The results do not only support a power-blocklength joint optimization of one-shot transmission scheme itself, but may also shed light on a future breakthrough to the CLARQ mechanism.

The remainder of this paper is organized as follows: Sec.~\ref{sec:setup} describes the system model and sets up the optimization problem, which is then analyzed in Sec.~\ref{sec:analysis}. In Sec.~\ref{sec:simulation} we numerically demonstrate our analyses, before closing the paper with concluding marks and outlooks in Sec.~\ref{sec:conclusion}.


\section{Problem Setup}\label{sec:setup}
\begin{figure}[!hbtp]
	\centering
	\includegraphics[width=.8\linewidth]{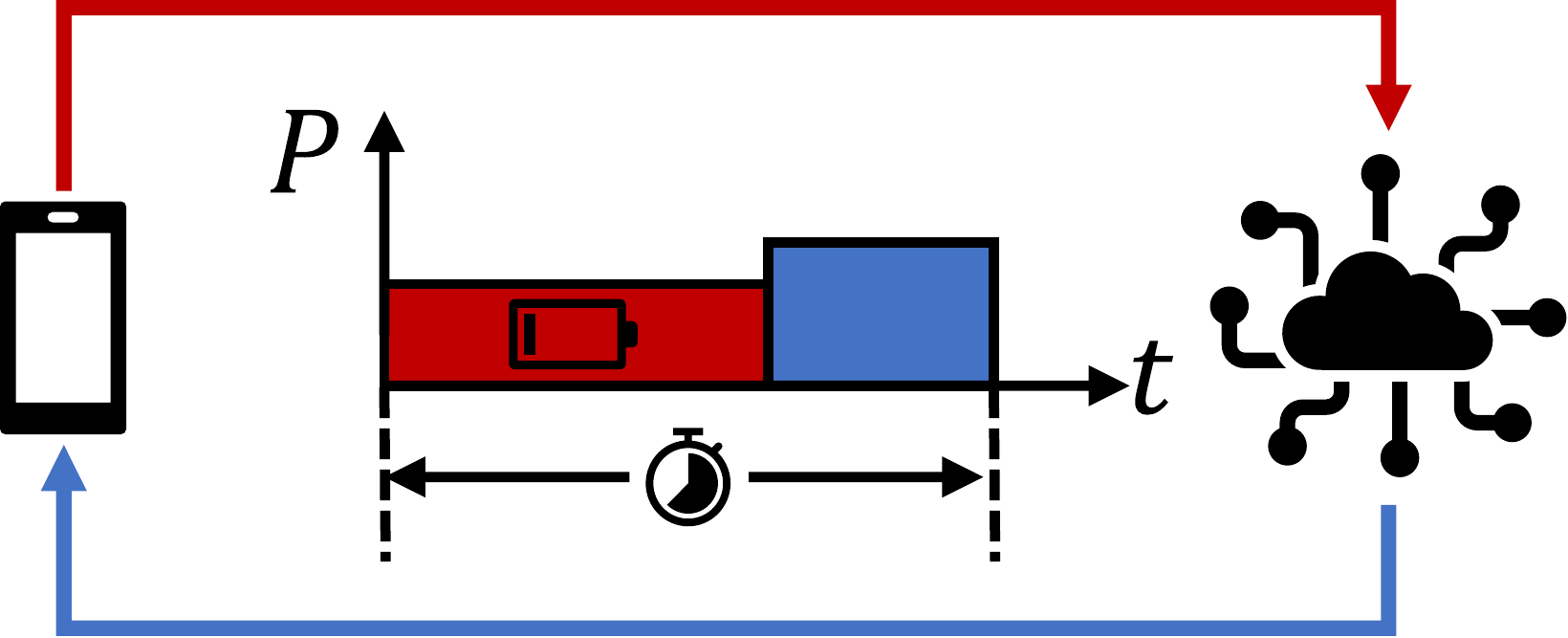}
	\caption{Model of the investigated system.}
	\label{fig:system_model}
\end{figure}

We consider a user equipment (UE) that communicates with the MEC server in a simple closed loop, as illustrated in Fig.~\ref{fig:system_model}. The UE periodically transmits in uplink (UL) to the server a message with $d$ bits of payload; upon a successful UL reception, the server responses in downlink (DL) also with $d$ bits of payload. The closed-loop air interface delay of the message exchange is limited to a certain frame length $T$. For the convenience of analysis, we omit here the processing time at server, and consider a certain sampling rate $f\subscript{s}$ with a fixed modulation order $2^M$ over a normalized channel bandwidth $B=\SI{1}{\hertz}$, so that the limited frame length is transferred into a limited blocklength $n\subscript{max}=f\subscript{s}T$ channel use\footnote{Can be normalized into $n\subscript{max}M$ bits}, which is flexibly shared by the UL and DL codewords of $n\ul$ and $n\dl$ channel use, respectively. We consider lossless channel coding in both UL and DL, i.e. $n\ul\geqslant d$ and $n\dl\geqslant d$. The UE has a fixed energy budget $E$ for each UL transmission, and is able to flexibly select its UL transmission power $p\ul$. 

For simplification of analysis we consider the background noise power $N$, the channel gains in UL $\vert h\subscript{UL}\vert^2$ and in DL $\vert h\subscript{DL}\vert^2$, as well as the DL transmission power $p\dl$ as all deterministic\footnote{In practical scenarios with block-fading channel conditions, the noise power and channel gains can be periodically measured, and the DL power adaptation without energy constraint can be straightforwardly addressed with off-the-shelf solutions, which is not the focus of this paper.}. In both link directions we consider short codes, so that the message error rate can be modeled as given by~\cite{polyanskiy2010channel}: 
\begin{equation}
	\varepsilon\subscript{L}=Q\left(\sqrt{\frac{n\subscript{L}}{V\subscript{L}}}\left(\mathcal{C}\subscript{L}-\frac{d}{n\subscript{L}}\right)\ln{2}\right),
\end{equation}
where $(\cdot)\subscript{L}\in\left\{(\cdot)\ul,(\cdot)\dl\right\}$, $\mathcal{C}\subscript{L}=B\log_2\left(1+\gamma\subscript{L}\right)$ is the Shannon's capacity, $V\subscript{L}=1-\frac{1}{\left(1+\gamma\subscript{L}\right)^{2}}$ the dispersion of AWGN channels, and $\gamma\subscript{L}=\frac{p\subscript{L}\vert h\subscript{L}\vert^2}{N}$ the SNR.

Aiming at a reliable MEC service under stringent latency requirement and energy limit, it calls for a joint optimization of the power and blocklength of UL transmission w.r.t. the closed-loop communication reliability:

\begin{maxi!}[2]
	{p\ul,n\ul}{R\subscript{loop}=(1-\varepsilon\ul)(1-\varepsilon\dl)}{\label{prob:main}}{}
	\addConstraint{n\ul+n\dl\leqslant n\subscript{max}\label{con:blocklength}}
	\addConstraint{\frac{n\ul p\ul}{f\subscript{s}}\leqslant E\label{con:energy_budget}}
	\addConstraint{\varepsilon\ul\leqslant\varepsilon\subscript{max},\varepsilon\dl\leqslant\varepsilon\subscript{max}\label{con:max_err}}
	\addConstraint{p\ul\in\mathbb{R}^+\label{con:pos_power}}
	\addConstraint{n\ul\in\mathbb{N},n\dl\in\mathbb{N}\label{con:integer_blocklength}}.
\end{maxi!}
Note that here we follow the routine approach of FBL information theory, setting an error rate bound $\varepsilon\subscript{max}$ for both link directions.

\section{Analysis}\label{sec:analysis}
\subsection{Approximations and Term Definitions}
The original problem \eqref{prob:main} is a multi-variate integer linear program (ILP). As commonly done in the FBL field, we relax the constraint of integer blocklength \eqref{con:integer_blocklength}, allowing to take real-valued blocklengths. Once a non-integer optimum is solved, the integer optimum can be obtained by rounding it~\cite{zhu2019reliability}. 

Besides, targeting at a high $R\subscript{loop}$, in practical scenarios it generally requires a stringently limited error rate in both UL and DL, so it is commonly approximated in the field of FBL that $\varepsilon\ul\varepsilon\dl\approx 0$, as done in \cite{hu2019swipt} and \cite{zhu2021error}. Thus, $R\subscript{loop}\approx1-\varepsilon\ul-\varepsilon\dl$. 

Additionally, since the interference control is out of the scope of this work, it naturally suggests to fully utilize the available blocklength and the energy budget, replacing the inequality constraints \eqref{con:blocklength} and \eqref{con:energy_budget} with their equality version. 

Thus, the original problem is transformed into a single-variate linear program (LP):
\begin{mini!}[2]
	{n\ul}{\varepsilon\cl\approx\varepsilon\ul+\varepsilon\dl}{\label{prob:main_lp}}{}
	\addConstraint{n\ul+n\dl= n\subscript{max}\label{con:blocklength_equality}}
	\addConstraint{\frac{n\ul p\ul}{f\subscript{s}}= E\label{con:energy_budget_equality}}
	\addConstraint{\varepsilon\ul\leqslant\varepsilon\subscript{max},\varepsilon\dl\leqslant\varepsilon\subscript{max}}
	\addConstraint{n\ul\in[d,n\subscript{max}-d]\label{con:real_blocklength}}.
\end{mini!}

Before any attempt to solve \eqref{prob:main_lp}, the existence of optimum must be confirmed, which is the main target of this paper. We invoke the sufficient KKT conditions, which requires to test the first-order and second-order derivatives of the objective function about $n\ul$.

\subsection{First-Order Derivative Test}
We start with the first-order derivative
\begin{equation}
	\frac{\partial\varepsilon\cl}{\partial n\ul}=\frac{\partial\varepsilon\ul}{\partial n\ul}+\frac{\partial\varepsilon\dl}{\partial n\ul}.
\end{equation}

For convenience of notation we define for both $(\cdot)\subscript{L}\in\{(\cdot)\ul,(\cdot)\dl\}$ the following terms:
\begin{align}
	\omega\subscript{L}&\triangleq \mathcal{C}\subscript{L}-\frac{d}{n\subscript{L}},\label{eq:omega}\\
	\beta\subscript{L}&\triangleq \sqrt{\frac{n\subscript{L}}{V\subscript{L}}},\label{eq:beta}\\
	x\subscript{L}&\triangleq (\ln2)\omega\subscript{L}\beta\subscript{L},\label{eq:x}\\
	\phi\subscript{L}&\triangleq-\frac{(\ln2)}{\sqrt{2\pi}}\exp\left(-\frac{x\subscript{L}^2}{2}\right)\label{eq:phi}.
\end{align}
Note that it always hold $\gamma\subscript{L}\geqslant0$ and $\phi\subscript{L}\leqslant0$. Therewith we have
\begin{align}
	\begin{split}\label{eq:diff_varepsilon_ul_0}
		\frac{\partial\varepsilon\ul}{\partial n\ul}
		=&\frac{\partial Q(x\ul)}{\partial n\ul}
		=-\frac{1}{\sqrt{2\pi}}\exp\left(-\frac{x\ul^2}{2}\right)\frac{\partial x\ul}{\partial n\ul}\\
		=&\phi\ul\left(\beta\ul\frac{\partial \omega\ul}{\partial n\ul}+\omega\ul\frac{\partial\beta\ul}{\partial n\ul}\right),
	\end{split}\\
	\begin{split}\label{eq:diff_varepsilon_dl_1}
		\frac{\partial\varepsilon\dl}{\partial n\ul}
		=&\frac{\partial Q(x\dl)}{\partial n\ul}
		=-\frac{1}{\sqrt{2\pi}}\exp\left(-\frac{x\dl^2}{2}\right)\frac{\partial x\dl}{\partial n\ul}\\
		=&\phi\dl\left(\beta\dl\frac{\partial \omega\dl}{\partial n\ul}+\omega\dl\frac{\partial\beta\dl}{\partial n\ul}\right).
	\end{split}
\end{align}

In the UL, with the fully utilized energy budget \eqref{con:energy_budget_equality}, we can derive the SNR as
\begin{equation}
	\gamma\ul=\frac{p\ul\vert h\ul\vert^2}{N}=\frac{Ef\subscript{s}\vert h\ul\vert^2}{Nn\ul},
\end{equation}
which leads to
\begin{align}
	\begin{split}\label{eq:diff_gamma_ul}
		\frac{\partial \gamma\ul}{\partial n\ul}
		=&\frac{\partial}{\partial n\ul}\frac{Ef\subscript{s}\vert h\ul\vert^2}{Nn\ul}
		=-\frac{Ef\subscript{s}\vert h\ul\vert^2}{Nn\ul^2}\\
		=&-\frac{\gamma\ul}{n\ul},
	\end{split}\\
	\begin{split}\label{eq:diff_omega_ul}
		\frac{\partial \omega\ul}{\partial n\ul}
		=&\frac{\partial\mathcal{C}\ul}{\partial n\ul}+\frac{d}{n\ul^2}
		=\frac{\partial\gamma\ul/\partial n\ul}{(\ln2)(1+\gamma\ul)}+\frac{d}{n\ul^2}\\
		=&-\frac{\gamma\ul}{(\ln2)(1+\gamma\ul)n\ul}+\frac{d}{n\ul^2}\\
		=&\frac{(\ln2)(1+\gamma\ul)d-\gamma\ul n\ul}{(\ln2)(1+\gamma\ul)n\ul^2},
	\end{split}
\end{align}
and
\begin{equation}
	\begin{split}\label{eq:diff_beta_ul}
		\frac{\partial\beta\ul}{\partial n\ul}
		=&\frac{1}{2\beta\ul}\frac{\partial}{\partial n\ul}\left(\frac{n\ul}{V\ul}\right)\\
		=&\frac{1}{2\beta\ul}\left(\frac{1}{V\ul}-\frac{n\ul}{V\ul^2}\frac{\partial V\ul}{\partial n\ul}\right)\\
		=&\frac{1}{2\beta\ul}\left[\frac{1}{V\ul}-\frac{2n\ul(\partial\gamma\ul/\partial n\ul)}{V\ul^2(1+\gamma\ul)^3}\right]\\
		=&\frac{1}{2\beta\ul}\left[\frac{1}{V\ul}+\frac{2\gamma\ul}{V\ul^2(1+\gamma\ul)^3}\right]\\
		=&\frac{V\ul (1+\gamma\ul)^3+2\gamma\ul}{2\beta\ul V\ul^2(1+\gamma\ul)^3}.
	\end{split}
\end{equation}
Thus, we have
\begin{equation}\label{eq:diff_varepsilon_ul}
	\begin{split}
		&\frac{\partial\varepsilon\ul}{\partial n\ul}\\
		=&\phi\ul\left[\beta\ul\frac{(\ln2)(1+\gamma\ul)d-\gamma\ul n\ul }{(\ln2)(1+\gamma\ul)n\ul^2}\right.\\
		&+\left.\omega\ul\frac{2\gamma\ul+V\ul (1+\gamma\ul)^3}{2\beta\ul V\ul^2(1+\gamma\ul)^3}\right]\\
		=&\frac{\phi\ul}{2(\ln2) \beta\ul V\ul^2(1+\gamma\ul)^3n\ul^2}\times\{2\beta\ul^2V\ul^2\\
		&\left.(1+\gamma\ul)^2\left[(\ln2)(1+\gamma\ul)d-\gamma\ul n\ul\right]\right.\\
		&\left.+(\ln2) \omega\ul n\ul^2\left[2\gamma\ul +V(1+\gamma\ul)^3\right]\right\}\\
		=&\underset{\triangleq\xi}{\underbrace{\frac{\phi\ul}{2(\ln2) \beta\ul V\ul^2n\ul(1+\gamma\ul)^3}}}\\
		&\times\left\{2V\ul(1+\gamma\ul)^2\left[(\ln2)(1+\gamma\ul)d-\gamma\ul n\ul\right]\right.\\		
		&\left.+(\ln2)\omega\ul n\ul\left[2\gamma\ul +V(1+\gamma\ul)^3\right]\right\}\\
		=&\xi\left\{((\ln2)\mathcal{C}\ul-1)V\ul\gamma\ul^3n\ul\right.\\
		&+(3(\ln2)\mathcal{C}\ul V\ul-2)\gamma\ul^2n\ul+[2(\ln2)\omega\ul n\ul\\
		&\left.+(3(\ln2)\mathcal{C}\ul-1)V\ul n\ul]\gamma\ul\right.\\
		&\left.+(\ln2)\mathcal{C}\ul V\ul n\ul\right\}
	\end{split}
\end{equation}
Now note that $\gamma\ul n\ul=\frac{p\ul n\ul\vert h\ul\vert^2}{N}=\frac{Ef\subscript{s}\vert h\ul\vert^2}{N}$ is a constant, which we let denoted by $\eta$. Recalling $V\ul=1-\frac{1}{(1+\gamma\ul)^2}=\frac{\gamma\ul^2+2\gamma\ul}{(1+\gamma\ul)^2}$, we have
\begin{equation}\label{eq:diff_varepsilon_ul_2}
	\begin{split}
		\frac{\partial\varepsilon\ul}{\partial n\ul}=&\xi\eta\left[\frac{((\ln2)\mathcal{C}\ul-1)(\gamma\ul^2+2\gamma)\gamma\ul^2}{(1+\gamma\ul)^2}\right.\\
		&+\frac{3(\ln2)\mathcal{C}\ul(\gamma\ul^2+2\gamma\ul-2)}{(1+\gamma\ul)^2}\\
		&+2(\ln2)\omega+\frac{(3(\ln2)\mathcal{C}\ul-1)(\gamma\ul^2+2\gamma\ul)}{(1+\gamma\ul)^2}\\
		&\left.+\frac{(\ln2)\mathcal{C}\ul(\gamma\ul+2)}{(1+\gamma\ul)^2}\right]\\
		=&\xi\eta\left[2(\ln2)\omega\ul+\delta\ul\right],
	\end{split}
\end{equation}
where
\begin{equation}
	\begin{split}
		\delta\ul\triangleq&\frac{1}{(1+\gamma\ul)^2}\times\left[(4(\ln2)\mathcal{C}\ul-3)\gamma\ul^3\right.\\
		&\left.+(11(\ln2)\mathcal{C}\ul-7)\gamma\ul^2+(\ln2)\mathcal{C}\ul\gamma\ul+2\right]
	\end{split}
\end{equation}

The sign of \eqref{eq:diff_varepsilon_ul_2} is not consistent over the entire feasible region. However, if we require an over-0-dB SNR in UL, i.e. $\gamma\ul\geqslant 1$, there will be
\begin{align}
	\mathcal{C}\ul\geqslant& 1\\
	\omega\ul\geqslant&0
\end{align}
Furthermore, we can also derive in this case that $\frac{\partial\delta\ul}{\partial\gamma\ul}>0$:
\begin{align}
	\mathcal{C}\ul\vert_{\gamma\ul=1}
	=&\log_2(1+1)=1\\
	\begin{split}
		\delta\ul\vert_{\gamma\ul=1}
		=&\frac{4(\ln2)-3+11(\ln2)-7+(\ln2)+2}{(1+1)^2}\\
		=&\frac{16(\ln2)-8}{4}>0
	\end{split}\\
	\begin{split}\label{eq:diff_delta_ul_to_gamma_ul}
		\frac{\partial\delta\ul}{\partial\gamma\ul}
		=&\frac{1}{(1+\gamma\ul)^3}\left[(4(\ln2)\mathcal{C}\ul+1)\gamma\ul^3\right.\\
		&+(12(\ln2)\mathcal{C}\ul+2)\gamma\ul^2\\
		&+(21(\ln2)\mathcal{C}\ul-13)\gamma\ul\\
		&\left.+((\ln2)\mathcal{C}\ul-4)\right]>0,\quad\forall\gamma\ul\geqslant 1
	\end{split}
\end{align}
Especially, the condition $\gamma\ul\geqslant 1$ is equivalent to $n\ul\leqslant\frac{Ef\subscript{s}\vert h\subscript{UL}\vert^2}{N}$ under the constraint \eqref{con:energy_budget_equality}. 
Hence, as $\xi<0$ and $\eta\geqslant0$, in the range $n\ul\in\left[d,\min\left(\frac{Ef\subscript{s}\vert h\subscript{UL}\vert^2}{N},n\subscript{max}-d\right)\right]$, it always holds $\frac{\partial\varepsilon\ul}{\partial n\ul}\leqslant0$.

On the other hand, for the DL, since 
\begin{align}
	\frac{\partial \gamma\dl}{\partial n\ul}=&0,\\
	\frac{\partial \omega\dl}{\partial n\ul}
	=&-\frac{d}{n\dl^2},\\
	\frac{\partial\beta\dl}{\partial n\ul}
	=&-\frac{1}{2\beta\dl V\dl},
\end{align}
obviously there is always
\begin{equation}\label{eq:diff_varepsilon_dl_2}
	\begin{split}
		\frac{\partial\varepsilon\dl}{\partial n\ul}=-\phi\dl\left(\frac{\beta\dl d}{n\dl^2}+\frac{\omega\dl}{2\beta\dl V\dl}\right)>0,\\
		\forall n\ul\in[d,n\subscript{max}-d]
	\end{split}
\end{equation}

Now we see that as $n\ul$ increases within the range $\left[d,\min\left(\frac{Ef\subscript{s}\vert h\subscript{UL}\vert^2}{N},n\subscript{max}-d\right)\right]$, $\varepsilon\ul$ monotonically decreases and $\varepsilon\dl$ monotonically increases. How their sum changes, however, highly depends on $n\subscript{max}$ and $p\subscript{DL}$.

\subsection{Second-Order Derivative Test}
Then we investigate the second-order derivative $\frac{\partial^2\varepsilon\cl}{\partial n\ul^2}=\frac{\partial^2\varepsilon\ul}{\partial n\ul^2}+\frac{\partial^2\varepsilon\dl}{\partial n\ul^2}$. Starting with UL:
\begin{equation}
	\begin{split}
		\frac{\partial^2\varepsilon\ul}{\partial n\ul^2}\overset{\eqref{eq:diff_varepsilon_ul}}{=}
		&\underset{X1}{\underbrace{\frac{\partial\xi}{\partial n\ul}\left[\eta(2(\ln 2)\omega\ul+\delta\ul)\right]}}\\
		&+\underset{X2}{\underbrace{\frac{\partial\eta}{\partial n\ul}[\xi(2(\ln2)\omega\ul+\delta\ul)]}}\\
		&+\underset{X3}{\underbrace{\xi\eta\left(2(\ln2)\frac{\partial\omega\ul}{\partial n\ul}+\frac{\partial\delta\ul}{\partial n\ul}\right)}}.
	\end{split}
\end{equation}
First we analyze the $X_1$ term, since
\begin{align}
	\begin{split}\label{eq:diff_phi_ul}
		\frac{\partial\phi\ul}{\partial n\ul}\overset{\eqref{eq:phi}}{=}&-\frac{\ln2}{\sqrt{2\pi}}\exp\left(-\frac{x\ul^2}{2}\right)(-x\ul)\frac{\partial x\ul}{\partial n\ul}\\
		\overset{\eqref{eq:x},\eqref{eq:phi}}{=}&-\phi\ul(\ln2)^2\beta\ul\frac{\partial\beta\ul}{\partial n\ul},
	\end{split}\\
	\frac{\partial V\ul}{\partial n\ul}=&\frac{2}{(1+\gamma\ul)^3}\frac{\partial\gamma\ul}{\partial n\ul}\overset{\eqref{eq:diff_gamma_ul}}{=}-\frac{2\gamma\ul}{(1+\gamma\ul)^3n\ul}\label{eq:diff_v_ul}
\end{align}
we can decompose
\begin{align}
		\frac{\partial\xi}{\partial n\ul}\overset{\xi\text{ in }\eqref{eq:diff_varepsilon_ul}}{=}&\frac{\partial\phi\ul/\partial n\ul}{2(\ln2)\beta\ul V\ul^2n\ul(1+\gamma\ul)^3}\\
		&+\phi\ul\left(\frac{\partial}{\partial n\ul}\frac{1}{2(\ln2)\beta\ul V\ul^2n\ul(1+\gamma\ul)^3}\right)\nonumber\\
		\overset{\eqref{eq:diff_phi_ul}}{=}&\frac{-\phi\ul(\ln2)^2\beta\ul(\partial\beta\ul/n\ul)}{2(\ln2)\beta\ul V\ul^2n\ul(1+\gamma\ul)^3}\nonumber\\
		&+\frac{\phi\ul}{2(\ln2)\beta\ul V\ul^2n\ul(1+\gamma\ul)^3}\nonumber\\
		&\times\left(-\frac{\partial\beta\ul/\partial n\ul}{\beta\ul}-\frac{2\partial V\ul/\partial n\ul}{V\ul}\right.\nonumber\\
		&\left.-\frac{1}{n}-\frac{3\partial\gamma\ul/\partial n\ul}{1+\gamma\ul}\right)\nonumber\\
		\overset{\xi\text{ in }\eqref{eq:diff_varepsilon_ul}}{=}&-\xi\left\{\left[(\ln2)^2\beta\ul+\frac{1}{\beta\ul}\right]\frac{\partial\beta\ul}{\partial n\ul}\right.\nonumber\\
		&\left.+\frac{2}{V\ul}\frac{\partial V\ul}{\partial n\ul}+\frac{1}{n\ul}+\frac{3}{1+\gamma\ul}\frac{\partial \gamma\ul}{\partial n\ul}\right\}\nonumber\\
		\overset{\eqref{eq:diff_beta_ul},\eqref{eq:diff_v_ul},\eqref{eq:diff_gamma_ul}}{=}&-\xi\left\{\left[(\ln2)^2\beta\ul+\frac{1}{\beta\ul}\right]\right.\nonumber\\
		&\times\frac{V\ul (1+\gamma\ul)^3+2\gamma\ul}{2\beta\ul V\ul^2(1+\gamma\ul)^3}\nonumber\\
		&-\frac{4\gamma\ul}{V\ul(1+\gamma\ul)^3n\ul}+\frac{1}{n\ul}\nonumber\\
		&\left.-\frac{3\gamma\ul}{(1+\gamma\ul)n\ul}\right\}\nonumber
\end{align}
\begin{align}				
		=&-\frac{\xi}{2V\ul(1+\gamma\ul)^3n\ul}\left\{[(\ln2)^2n\ul\right.\nonumber\\
		&+V\ul][V\ul(1+\gamma\ul)^3+2\gamma\ul]-8\gamma\ul\nonumber\\
		&\left.+2V\ul(1+\gamma\ul)^3-6V\ul\gamma\ul(1+\gamma\ul)^2\right\}\nonumber.
\end{align}
For convenience we define $\rho\triangleq(\ln2)^2n\ul$, so the equation above can be rearranged into
\begin{equation}
	\begin{split}
		\frac{\partial\xi}{\partial n\ul}=&-\frac{\xi}{2V\ul(1+\gamma\ul)^3n\ul}\left\{(\rho+V\ul)\right.\\
		&\times[V\ul(1+\gamma\ul)^3+2\gamma\ul]-8\gamma\ul\\
		&\left.+2V\ul(1+\gamma\ul)^3-6V\ul\gamma\ul(1+\gamma\ul)^2\right\}\\
		=&-\frac{\xi}{2V\ul(1+\gamma\ul)^3n\ul}\times[(\rho -4)V\ul\gamma\ul^3\\
		&+(3\rho -6)V\ul\gamma\ul^2+(3\rho V\ul+2\rho-8)\gamma\ul\\
		&+\rho+2V\ul].
	\end{split}
\end{equation}
Here we introduce a new assumption that $n\ul\geqslant 9$, which is a loose constraint for most practical systems. Under this condition, we have $\rho\geqslant 9(\ln2)^2>4$. Since $\xi<0$, this guarantees
\begin{equation}
	\frac{\partial\xi}{\partial n\ul}>0.
\end{equation}
Besides, it also always holds that $\eta>0$. Thus, with $\gamma\ul\geqslant1$ and $n\geqslant d$ that force $\omega\ul\geqslant0$, we know $X_1>0$.

The term $X_2$ is obviously zero, since $\frac{\partial\eta}{\partial n\ul}=0$.

For term $X_3$, we know $\xi<0$, $\eta>0$, and
\begin{align}
		&2(\ln2)\frac{\partial\omega\ul}{\partial n\ul}+\frac{\partial\delta\ul}{\partial n\ul}\nonumber\\
		\overset{\eqref{eq:diff_omega_ul}}{=}&\frac{2\left[(\ln2)(1+\gamma\ul)d-\gamma\ul n\ul\right]}{(1+\gamma\ul)n\ul^2}+\frac{\partial\delta\ul}{\partial\gamma\ul}\frac{\partial\gamma\ul}{\partial n\ul}\nonumber\\
		\overset{\eqref{eq:diff_gamma_ul},\eqref{eq:diff_delta_ul_to_gamma_ul}}{=}&\frac{2\left[(\ln2)(1+\gamma\ul)d-\gamma\ul n\ul\right]}{(1+\gamma\ul)n\ul^2}-\frac{\gamma\ul}{(1+\gamma\ul)^3n\ul}\nonumber\\
		&\times\left[(4(\ln2)\mathcal{C}\ul+1)\gamma\ul^3+(12(\ln2)\mathcal{C}\ul+2)\gamma\ul^2\right.\nonumber\\
		&\left.+(21(\ln2)\mathcal{C}\ul-13)\gamma\ul+((\ln2)\mathcal{C}\ul-4)\right]\nonumber\\
		=&\frac{1}{(1+\gamma\ul)^3n\ul^2}\times[\underset{< -3.77n\ul,\forall\gamma\ul\geqslant1}{\underbrace{-(4(\ln2)\mathcal{C}\ul+1)n\ul\gamma\ul^4}}\nonumber\\
		&+\underset{< -3.3n\ul,\forall(\gamma\ul\geqslant1,n\ul\geqslant d)}{\underbrace{(2(\ln2)d-4n\ul-(\ln2)\mathcal{C}\ul n\ul)\gamma\ul^3}}\\
		&+\underset{< -2.78n\ul,\forall(\gamma\ul\geqslant1,n\ul\geqslant d)}{\underbrace{(4(\ln2)d-21(\ln2)\mathcal{C}\ul n\ul+9n\ul)\gamma\ul^2}}\nonumber\\
		&+\underset{<4.08n\ul,\forall(\gamma\ul\geqslant1,n\ul\geqslant d)}{\underbrace{(4(\ln2)d+2n\ul -(\ln2)\mathcal{C}n\ul)\gamma\ul}}\nonumber\\
		&+\underset{< 1.39n\ul,\forall n\ul\geqslant d}{\underbrace{2(\ln2)d}}]<0,\forall (\gamma\ul\geqslant1,n\ul\geqslant d),\nonumber
\end{align}
so that $X_3>0$.

Summing three terms up, we have
\begin{equation}
	\frac{\partial^2\varepsilon\ul}{\partial n\ul^2}>0,\quad\forall n\in\left[\max(9,d),\frac{Ef\subscript{s}\vert h\ul\vert^2}{N}\right]
\end{equation}

On the other hand, for the DL:
\begin{align}
		\frac{\partial^2\varepsilon\dl}{\partial n\ul^2}=&-\frac{\partial\phi\dl}{\partial n\ul}\left(\frac{\beta\dl d}{n\dl^2}+\frac{\omega\dl}{2\beta\dl V\dl}\right)\\
		&-\phi\dl\frac{\partial}{\partial n\ul}\left(\frac{\beta\dl d}{n\dl^2}+\frac{\omega\dl}{2\beta\dl V\dl}\right)\nonumber\\
		=&-\frac{\ln 2\phi\dl x\dl}{2\beta\dl V\dl}\left(\frac{\beta\dl d}{n\dl^2}+\frac{\omega\dl}{2\beta\dl V\dl}\right)\nonumber\\
		&-\phi\dl\left[\frac{3d\beta\dl}{2n\dl^3}-\frac{d}{2\beta\dl V\dl n\dl^2}+\frac{\omega\dl}{4\beta\dl^3V\dl^2}\right]\nonumber
\end{align}
Note that $\phi\dl<0$, $\beta\dl>0$,$x\dl>0$, $n\dl>0$, $\mathcal{C}\dl>0$, $\beta\dl^2V\dl=n\dl$, so we have
\begin{align}
	&\frac{\beta\dl d}{n\dl^2}+\frac{\omega\dl}{2\beta\dl V\dl}=\frac{2\beta\dl^2V\dl d}{2\beta\dl V\dl n\dl^2}+\frac{\mathcal{C}\dl n-d}{2\beta\dl V\dl n\dl}\nonumber\\
	=&\frac{2d+\mathcal{C}\dl}{2\beta\dl V\dl n\dl}>0.
\end{align}
Furthermore,
\begin{equation}
	\begin{split}
		&\frac{3d\beta\dl}{2n\dl^3}-\frac{d}{2\beta\dl V\dl n\dl^2}+\frac{\omega\dl}{4\beta\dl^3V\dl^2}\\
		=&\frac{3d\beta\dl^2V\dl}{2\beta\dl V\dl n\dl^3}-\frac{d}{2\beta\dl V\dl n\dl^2}+\frac{\mathcal{C}\dl-d/n\dl}{4\beta\dl V\dl n\dl}\\
		=&\frac{1}{2\beta\dl V\dl n\dl^2}\left(\frac{3dn\dl}{n\dl}-d+\frac{\mathcal{C}\dl n\dl-d}{2}\right)\\
		=&\frac{1}{2\beta\dl V\dl n\dl^2}\left(\frac{\mathcal{C}\dl n\dl+3d}{2}\right)>0.
	\end{split}
\end{equation}
Thus, there is always
\begin{equation}
	\frac{\partial^2\varepsilon\dl}{\partial n\ul^2}>0,\quad\forall n\ul\in\mathbb{R}^+.
\end{equation}

\subsection{Unique Optimum}
So far we have seen that in the domain $n\ul\in\left[\max(9,d),\min\left(\frac{Ef\subscript{s}\vert h\ul\vert^2}{N},n\subscript{max}-d\right)\right]$, both $\varepsilon\ul$ and $\varepsilon\dl$ are monotonic convex functions of $n\ul$, decreasing and increasing, respectively. Thus, it is straightforward to see that $\varepsilon\ul+\varepsilon\dl$ is also a convex function of $n\ul$ in the referred domain, and an unique minimum of \eqref{prob:main_lp} exists. To identify the optimum, values of the first-order derivative $\frac{\partial\varepsilon\cl}{\partial n\ul}$
at the left bound $n\superscript{min}\ul=\max(9,d)$ and the right bound $n\superscript{max}\ul=\min\left(\frac{Ef\subscript{s}\vert h\ul\vert^2}{N},n\subscript{max}-d\right)$ shall be checked:
\begin{equation}
	\begin{split}
		n\ul\superscript{opt}=&\arg\min\limits_{n\ul}(\varepsilon\ul+\varepsilon\dl)\\
		=&\begin{cases}
			n\ul\superscript{min}\quad&\frac{\partial\varepsilon\cl}{\partial n\ul}\vert_{n\ul=n\ul\superscript{min}}\geqslant 0,\\
			n\ul\superscript{max}\quad&\frac{\partial\varepsilon\cl}{\partial n\ul}\vert_{n\ul=n\ul\superscript{max}}\leqslant 0,\\
			n:\frac{\partial\varepsilon\cl}{\partial n\ul}\vert_{n\ul=n}=0\quad&\text{otherwise}.
		\end{cases}
	\end{split}
\end{equation}
The first two cases that indicate a monotonicity of $\varepsilon\cl$ can be simply identified through numerical test regarding \eqref{eq:diff_varepsilon_ul_2} and \eqref{eq:diff_varepsilon_dl_2}, respectively.
In the last case, unfortunately, the solution of $\frac{\partial\varepsilon\cl}{\partial n\ul}=0$ appears intractable, and no closed-form expression can be obtained for the analytical solution. Nevertheless, with the sufficient KKT conditions derived in this paper, conventional iterative solvers can be applied for fast numerical solution.

\section{Numerical Verification}\label{sec:simulation}
\begin{figure}[!hbtp]
	\centering
	\includegraphics[width=\linewidth]{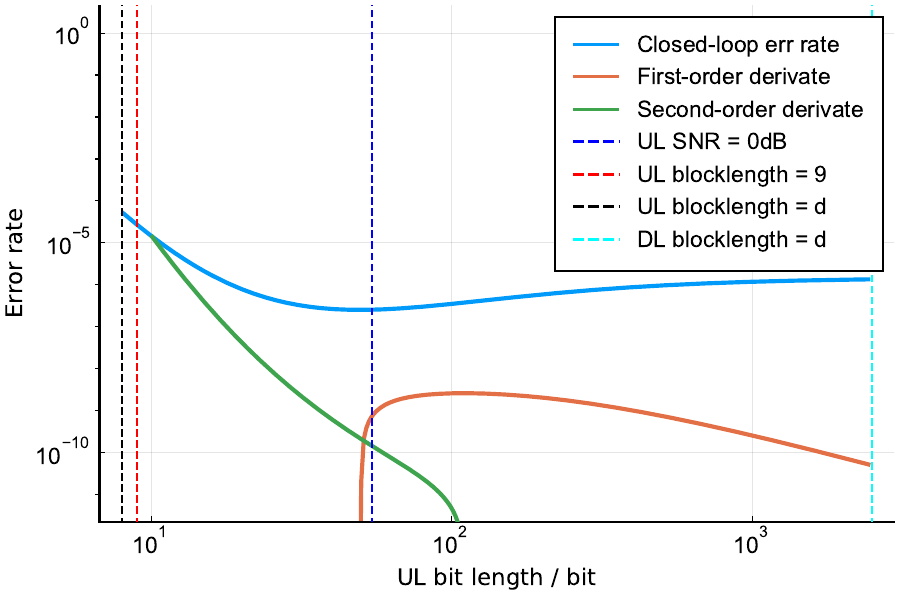}
	\caption{A case study of $\varepsilon\cl$ where $N=\SI{3}{\milli\watt}$.}
	\label{fig:case_study}
\end{figure}

To verify our analyses we conducted numerical simulations, for which the system was specified as listed in Tab.~\ref{tab:setup}.  Setting the background noise power $N=\SI{3}{\milli\watt}$ for a case study, we observed the closed-loop error rate $\varepsilon\cl$ and its derivatives within the key region of $n\ul$ constrained by the aforementioned key bounds. The exhibited behavior of $\varepsilon\cl$ matches our analysis well, as illustrated in Fig.\ref{fig:case_study}. It also reveals us that the $\SI{0}{\dB}$ UL SNR bound it much tighter than the $n\subscript{max}-d$ UL blocklength bound.

\begin{figure}[!hbtp]
	\centering
	\includegraphics[width=\linewidth]{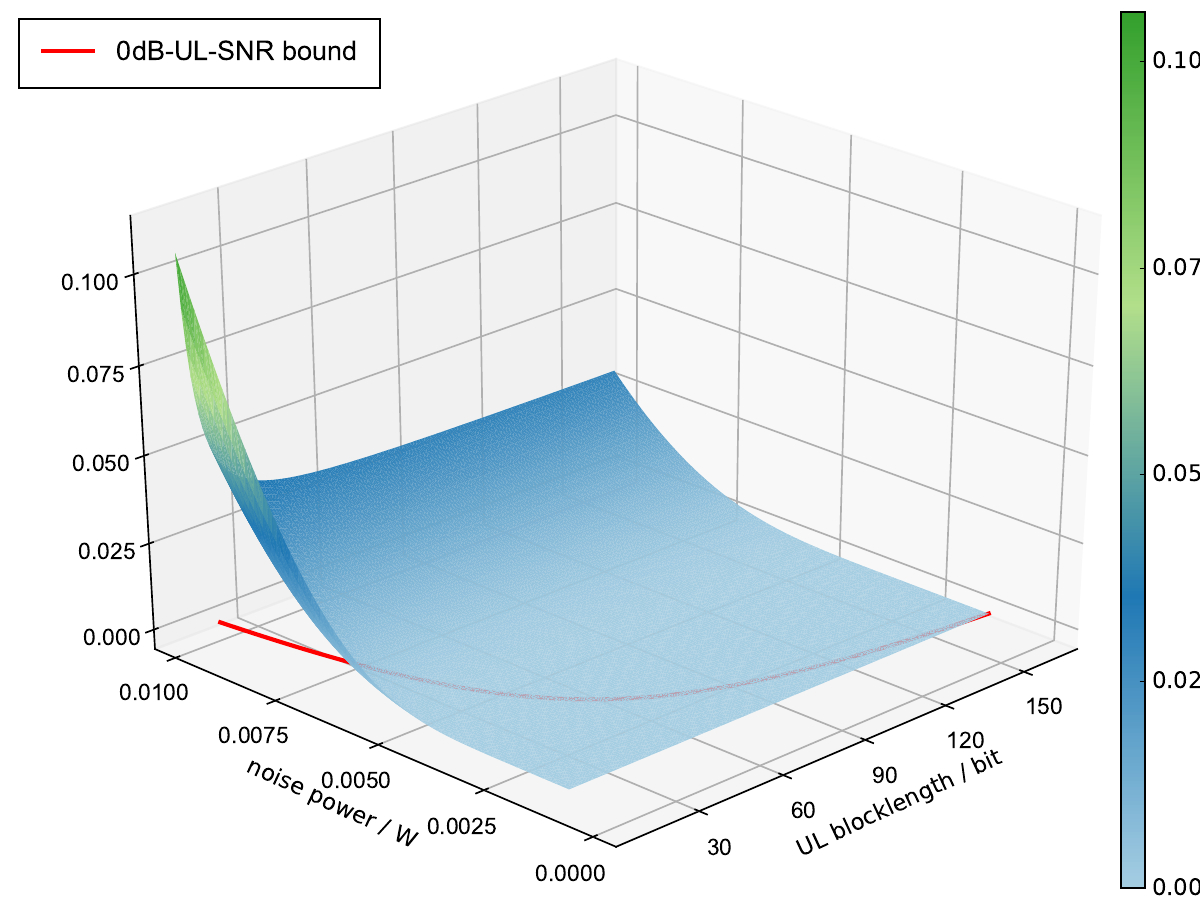}
	\caption{The closed-loop error rate $\varepsilon\cl$.}
	\label{fig:cler}
\end{figure}

\begin{figure}[!hbtp]
	\centering
	\includegraphics[width=\linewidth]{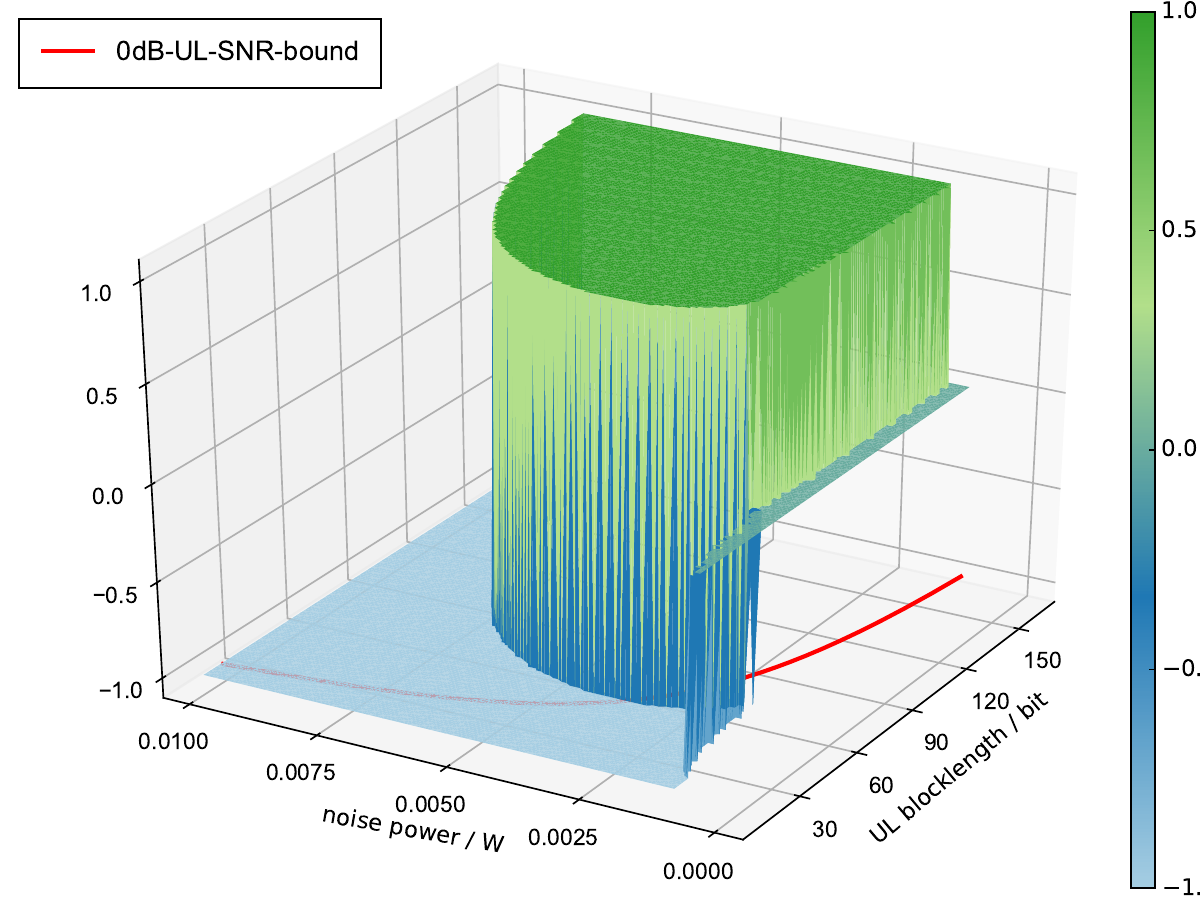}
	\caption{The sign of $\partial\varepsilon\cl/\partial n\ul$.}
	\label{fig:sign_derivative}
\end{figure}

\begin{figure}[!hbtp]
	\centering
	\includegraphics[width=\linewidth]{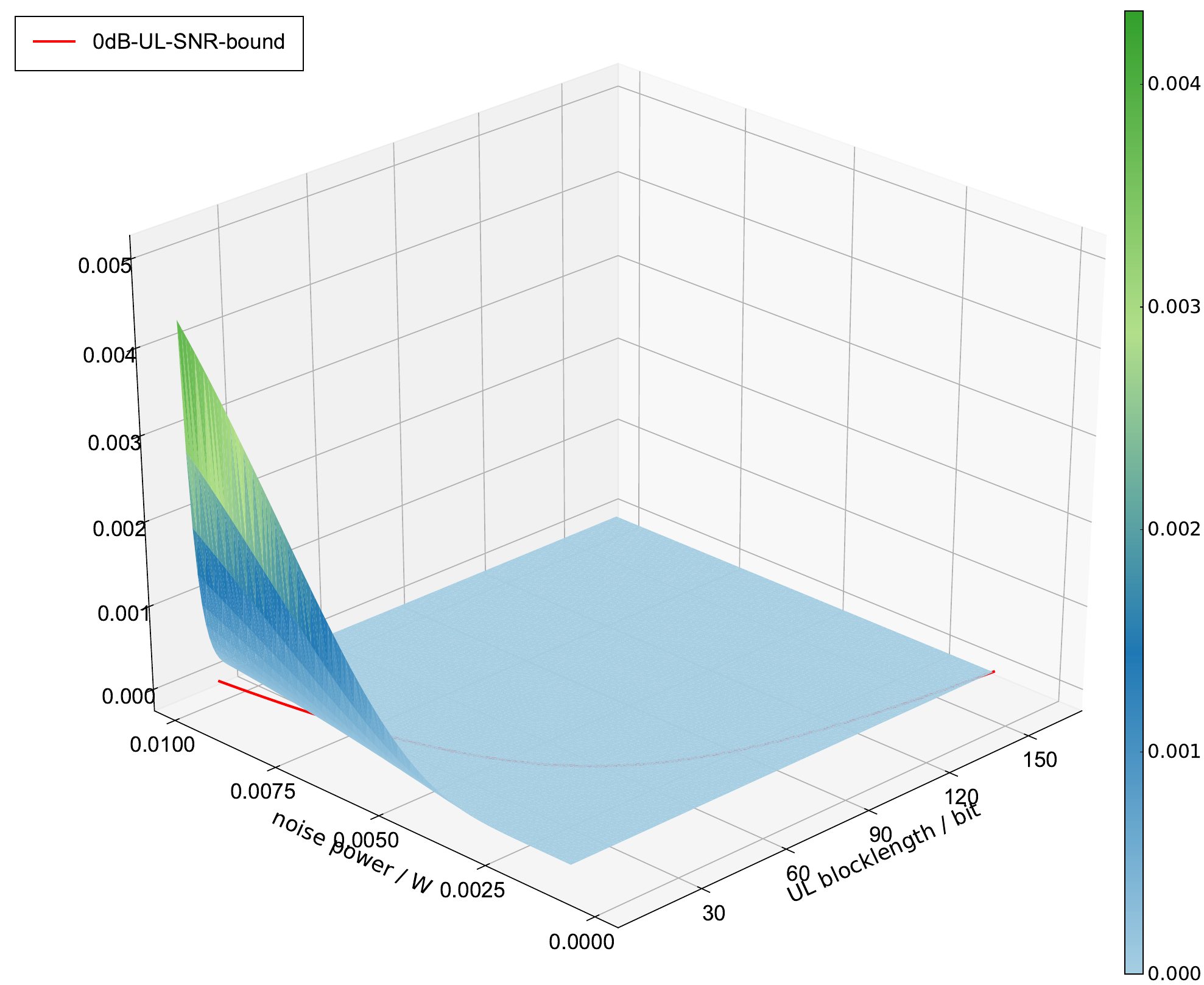}
	\caption{The second-order derivative $\partial^2\varepsilon\cl/\partial n\ul^2$.}
	\label{fig:second_derivative}
\end{figure}

To ensure the generality, we tested $\varepsilon\cl$, $\mathrm{sign}\left(\frac{\partial\varepsilon\cl}{\partial n\ul}\right)$ and $\frac{\partial^2\varepsilon\cl}{\partial n\ul^2}$ over the noise power range $N\in(0,p\dl)$, with the results depicted in Figs.~\ref{fig:cler}--\ref{fig:second_derivative}, respectively. From the figures we can confirm our analysis in Sec.~\ref{sec:analysis}, that $\varepsilon\cl$ \emph{is not guaranteed monotonic but convex} over $n\ul\in\left[\max(9,d),\min\left(\frac{Ef\subscript{s}\vert h\ul\vert^2}{N},n\subscript{max}-d\right)\right]$.

\begin{table}[!htbp]
	\centering
	\caption{Simulation setup}
	\label{tab:setup}
	\begin{tabular}{c|c|c}
		\toprule[2px]
		\textbf{Parameter}	&\textbf{Value}			&\textbf{Note}\\
		\midrule[1.5px]
		$f\subscript{s}$	&250kSPS				&\SI{4}{\micro\second} OFDM symbol duration (5G numerology 4)\\
		$M$					&1						&BPSK\\
		$n\subscript{max}$	&2500					&Corresponding to \SI{10}{\milli\second} latency w.r.t. $f\subscript{s}$ and $M$\\
		$\vert h\ul\vert^2$	&1						&\\
		$\vert h\dl\vert^2$	&1						&\\
		$p\dl$				&\SI{10}{\milli\watt}	&\\
		$d$					&8 bits					&\\
		\multirow{2}{*}{$E$}&\multirow{2}{*}{\SI{0.65}{\micro\joule}}
													&Allows a standard \SI{3.8}{\volt} / 3000 mAh battery\\
							&						& to work 10 years under 50\% UL duty rate\\
		\bottomrule[2px]
	\end{tabular}
\end{table}

\section{Conclusion and Outlooks}\label{sec:conclusion}
In this paper, we have studied the blocklength allocation problem regarding closed-loop reliability optimization in FBL regime, under stringent constraints of closed-loop latency and uplink energy. We have identified under weak assumptions a tight domain  where an unique optimum of this problem exists, and provided the characterizing features of the solution. The optimal blocklength allocation for the one-shot transmission scheme can be therewith obtained.

For future work, we are looking forward to extending the current results into the CLARQ case, in order to enable an optimal power control for the dynamic closed-loop ARQ mechanism.

\section*{Acknowledgment}
This research was supported by the German Federal Ministry of Education and Research (BMBF) within the project Open6GHub under grant numbers 16KISK003K, 16KISK004 and 16KISK012.


%

%
%

\ifCLASSOPTIONcaptionsoff
  \newpage
\fi

\begin{IEEEbiographynophoto}{Jane Doe}
Biography text here.
\end{IEEEbiographynophoto}




\end{document}